\documentclass{aa}  

\usepackage{graphicx}
\usepackage{amsmath,amssymb}
\usepackage{longtable,lscape}
\usepackage{txfonts}
\begin{document}

   \title{Dynamics, CO depletion, and deuterium fractionation of
the dense condensations within the fragmented prestellar core 
Orion B9--SMM 6\thanks{This publication is based on data acquired with the 
Atacama Pathfinder EXperiment (APEX) under programmes 
079.F-9313(A), 084.F-9312(A), and 090.F-9313(A). APEX is a collaboration 
between the Max-Planck-Institut f\"{u}r Radioastronomie, the European Southern 
Observatory, and the Onsala Space Observatory.}}


   \author{O. Miettinen\inst{1} and S.~S.~R. Offner\inst{2}\thanks{Hubble Postdoctoral Fellow.}}

   \institute{Department of Physics, University of Helsinki, P.O. Box 64, FI-00014 Helsinki, Finland\\ \email{oskari.miettinen@helsinki.fi} \and Department of 
Astronomy, Yale University, New Haven, CT 06511, USA \\ \email{stella.offner@yale.edu}}

   \date{Received ; accepted}

\authorrunning{Miettinen \& Offner}
\titlerunning{APEX observations of the fragmented prestellar core Ori B9--SMM 6}

  \abstract
   {Low-mass prestellar cores are rarely found to be fragmented into smaller 
condensations but studying such substructure where present is 
essential for understanding the origin of multiple stellar systems.}
   {We attempt to better understand the kinematics and dynamics of the 
subfragments inside the prestellar core SMM 6 in Orion B9. Another object of 
the present study is to constrain the evolutionary stage of the condensations 
by investigating the levels of CO depletion and deuterium fractionation.}
   {We used the APEX telescope to observe the molecular lines 
C$^{17}$O$(2-1)$, N$_2$H$^+(3-2)$, and N$_2$D$^+(3-2)$ towards the 
condensations. We use the line data in conjunction with our previous 
SABOCA 350-$\mu$m dust continuum map of the source.}
   {The condensations are characterised by subsonic internal non-thermal 
motions ($\sigma_{\rm NT}\simeq 0.5c_{\rm s}$), and most of them appear to be 
gravitationally bound. The dispersion of the N$_2$H$^+$ velocity 
centroids among the condensations is very low (0.02 km~s$^{-1}$). 
The CO depletion factors we derive, $f_{\rm D}=0.8\pm0.4-3.6\pm1.5$, 
do not suggest any significant CO freeze-out but this may be due to 
the canonical CO abundance we adopt. The fractional abundances of N$_2$H$^+$ 
and N$_2$D$^+$ with respect to H$_2$ are found to be 
$\sim0.9-2.3\times10^{-9}$ and $\sim4.9-9.9\times10^{-10}$, 
respectively. The deuterium fractionation of N$_2$H$^+$ lies
in the range $0.30\pm0.07-0.43\pm0.09$.}
   {The detected substructure inside SMM 6 is likely the result of cylindrical 
Jeans-type gravitational fragmentation. We estimate the timescale for 
this fragmentation to be $\sim1.8\times10^5$ yr. The condensations are 
unlikely to be able to interact with one another and coalesce before local 
gravitational collapse ensues. Moreover, significant mass growth of the 
condensations via competitive-like accretion from the parent core seems 
unfeasible. The high level of molecular deuteration in the condensations 
suggests that gas-phase CO should be strongly depleted. It also points towards 
an advanced stage of chemical evolution. The subfragments of SMM 6 might 
therefore be near the onset of gravitational collapse, but whether they can 
form protostellar or substellar objects (brown dwarfs) depends on the local 
star formation efficiency and remains to be clarified.}


   \maketitle
%

\section{Introduction}

Fragmentation of interstellar molecular clouds appers to occur in a 
hierarchical fashion. Recent \textit{Herschel} observations have 
clearly demonstrated that on the scales of several parsecs to $\sim10$ pc the 
cloud structures are highly filamentary in shape (e.g., \cite{andre2010}; 
\cite{arzoumanian2011}; \cite{hill2011}; \cite{palmeirim2013}). The origin 
of large-scale filaments is still unclear, but they often (if not always) show 
substructures along their long axes. Substructures on scales of 
$\sim0.5-1$ pc are typically called \textit{clumps}, which may contain 
significant gas substructure and would likely form a cluster of stars. Gas 
fragments $\lesssim0.1$ pc, which are expected to form a single star or bound 
multiple system, are called \textit{cores}.

The dense cores formed in the course of hierarchical 
fragmentation represent the size scale on which protostellar collapse
and the formation of low-mass ($\sim0.1-2$ M$_{\sun}$) stars take
place. Initially starless cores that are bound by their self-gravity 
are called \textit{prestellar cores} (\cite{wardthompson1994}). 
These are ideal laboratories to study the 
genuine initial physical and chemical conditions of star formation.
Because many stars are found in binary or higher-order multiple 
systems (e.g., \cite{raghavan2010}; \cite{kraus2012}; \cite{duchene2013}), 
core fragmentation into still smaller subunits is expected to occur at some 
point in their evolution, possibly during the prestellar phase. 
In the present work, we will use the term \textit{condensation} for such 
subfragments with sizes of $\sim0.01$ pc (e.g., \cite{andre2007}, hereafter 
ABMP07). Since condensations are fairly low-mass and are usually found within a 
gravitationally bound core, they likely represent the precursors of individual 
stars that will eventually comprise a multiple star system.

Core fragmentation is of prime importance when studying the origin of 
core mass function and its connection to the stellar initial mass 
function (IMF). From a theoretical point of view, the formation of multiple 
stellar systems is far from well-understood 
(e.g., \cite{tohline2002}; \cite{goodwin2007}). 
For example, the initial phase and characteristic timescale of the 
fragmentation process remain to be established. Observationally, it 
is unclear how common the fragmentation of low-mass cores actually is, 
particularly in the case of starless/prestellar cores. 

Some of the earliest instances of core substructure were discovered by 
Lemme et al. (1995). They mapped the starless core L1498 in the C-bearing 
molecules C$^{18}$O and CS using the IRAM 30-m telescope with 
$12\arcsec-25\arcsec$ resolution. The core was resolved into many small-scale 
condensations with diameters and masses of about 0.02 pc and 0.01 M$_{\sun}$, 
respectively. The subfragments were found to be gravitationally unbound, 
and the authors suggested that they are just transient structures 
(see also \cite{kuiper1996}). Langer et al. (1995) used high-resolution 
($6\arcsec-9\arcsec$) interferometric observations to map the 
TMC-1 core D. Similiarly to L1498, the core was found to be composed of 
several condensations with sizes and masses of 0.007--0.021 pc and $\sim0.01$ 
M$_{\sun}$, respectively. These sources were also found to be 
gravitationally unbound and unable to form even proto-brown dwarfs unless the 
condensations were to merge and form more massive structures. 
Langer et al. (1995) suggested that a pure Jeans-type gravitational 
instability could not be responsible for the core fragmentation. 
However, the substructure was identified from molecular-line maps 
(the C-bearing species CCS and CS). Therefore, it is possible that the 
subfragments represent small-scale chemical inhomogeneities of the parent 
core (cf.~\cite{takakuwa1998}).

Kamazaki et al. (2001) detected the first subfragments inside 
prestellar cores in dust emission. They carried out high-resolution NMA 
observations of 2- and 3-mm dust continuum emission towards two prestellar 
cores in the $\rho$ Oph A region, namely SM1 and SM1N. Both cores appeared to 
contain condensations of 600--1\,100 AU in size with masses in the range 
0.054--0.14 M$_{\sun}$. Interestingly, as opposed to the above mentioned 
studies, the condensations found by Kamazaki et al. (2001) were determined to 
be gravitationally bound entities. We note that one of the condensations 
within SM1, with a mass of 0.14 M$_{\sun}$, was suggested by the authors to be 
already in the early protostellar phase. Kamazaki et al. (2001) also 
concluded that the substructures were the result of 
gravitational fragmentation of the parent cores. Takakuwa et al. (2003) 
studied two prestellar cores in the TMC-1C region with high-resolution 
molecular-line observations, and found $\sim2\,500$ AU -size condensations 
with masses of $\sim0.02$ M$_{\sun}$ inside the cores. These substellar-mass 
fragments were found to be gravitationally unbound but with dissipation 
timescales long enough to allow the fragments to coalesce with one another 
and form more massive objects.

More recently, Kirk et al. (2009) studied the prestellar core L183 (L134N) with 
high-resolution ($13\arcsec \times 10\arcsec$ or $\sim0.006$ pc $\sim1\,200$ 
AU) BIMA interferometric observations, and found it to be composed of three 
condensations of $\sim0.005-0.01$ pc in radii. They concluded that the 
observed substructure was the result of the fragmentation of 
a rotating and collapsing prestellar core. On the other hand, Schnee et al. 
(2010) found that none of their 11 starless cores in Perseus break into 
smaller components at $5\arcsec$ resolution or on the $\sim10^3$ AU scale (see 
also \cite{miettinen2013} for Ori B9--SMM 1). Chen \& Arce (2010), using SMA 
observations at $5\farcs7 \times 2\farcs3$ resolution ($\sim0.003$ pc or 
$\sim620$ AU), discovered three condensations inside the prestellar core R CrA 
SMM 1A. The projected separation between the 0.1--0.2 M$_{\sun}$ mass 
condensations was found to be between 1\,000 and 2\,100 AU. 
The observed spacings are comparable to the local Jeans length, and the authors 
suggested that the condensations were formed through the fragmentation of an 
elongated prestellar core during the isothermal phase of evolution (i.e., when 
the radiative cooling is able to compensate for the heating by 
compression). Nakamura et al. (2012) used the SMA with a few arcsec (a few 
hundred AU) resolution to observe the prestellar cores SM1 and B2-N5 in 
$\rho$ Ophiuchus (cf.~\cite{kamazaki2001}). These cores were resolved into 
three and four condensations, respectively, with masses of $\sim0.01-0.1$ 
M$_{\sun}$ and sizes of a few hundred AU. The masses were determined to be high 
enough for the condensations to be gravitationally bound. 
Nakamura et al. (2012) proposed that the origin of the substructures 
could be explained by turbulent fragmentation (e.g., 
\cite{fisher2004}; \cite{offner2010}).

\begin{figure}[!h]
\centering
\resizebox{\hsize}{!}{\includegraphics{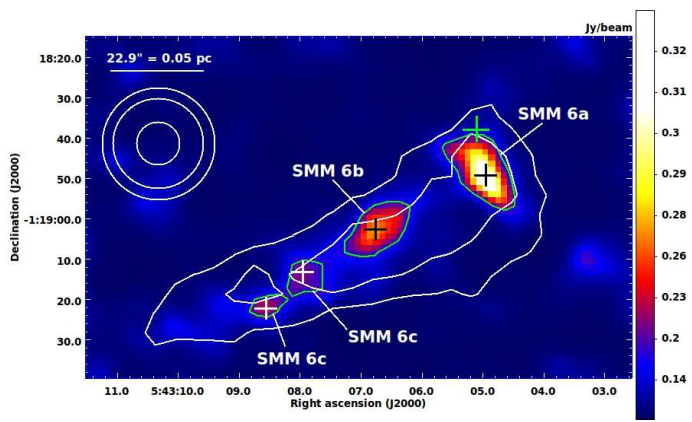}}
\caption{SABOCA 350-$\mu$m image of the fragmented prestellar
core SMM 6 in Orion B9. The image is displayed with power-law scaling, and the 
colour bar indicates the surface-brightness scale in Jy~beam$^{-1}$. The 
overlaid green SABOCA contours are plotted at three times the noise level 
($3\sigma=180$ mJy~beam$^{-1}$). The white contours show the 
\textit{Herschel}/SPIRE 250-$\mu$m dust continuum emission ($18\arcsec$ 
resolution); these contours are plotted at 3.0 and 3.5 Jy~beam$^{-1}$. 
The green plus sign indicates the target position of our previous 
molecular-line observations, while the other plus signs mark the 
target positions of the present study (i.e., the 350-$\mu$m peaks). 
A scale bar indicating the 0.05 pc projected 
length is shown in the top left, with the assumption of a 450 pc line-of-sight 
distance. The three circles in the upper left corner show 
the effective FWHM of the SABOCA beam ($10\farcs6$), and the smallest 
($22\farcs3$) and largest ($27\farcs8$) beamsizes of the present
molecular-line observations.}
\label{figure:SMM6}
\end{figure}

The target source of the present study is the prestellar core SMM 6 in the 
Orion B9 star-forming region. It was originally discovered by Miettinen et al. 
(2009; hereafter, Paper I) from the LABOCA 870-$\mu$m mapping of Orion B9. 
The core was found to be elongated in shape and not harbouring any 
embedded infrared point source, and we suggested that it is 
likely to be prestellar due its high density. Based on the NH$_3$ observations 
with the Effelsberg 100-m telescope, Miettinen et al. (2010; henceforth, 
Paper II) derived the gas kinetic temperature of $T_{\rm kin}=11.0\pm0.4$ K 
towards a selected position in SMM 6. 
The one-dimensional non-thermal velocity dispersion was derived to 
be 0.1 km~s$^{-1}$, which is subsonic by a factor of two. 
Employing the new temperature value with the assumption that 
$T_{\rm kin}$ equals the dust temperature ($T_{\rm dust}$), the mass of SMM 6 was 
estimated to be $8.2\pm1.1$ M$_{\odot}$. 
The virial-parameter analysis of the source suggested that it is 
gravitationally bound and near virial equilibrium, 
supporting our earlier speculation that it is in the prestellar phase of 
evolution.\footnote{The core mass is revised down by a factor of 
$\sim1.4$ in the present paper. This does not affect its classification 
into a gravitational bound or prestellar core (Appendix~A).} 
Miettinen et al. (2012; hereafter, Paper III) presented further molecular-line 
observations towards the NH$_3$ target position in SMM 6.
The CO depletion factor was estimated to be $4.2\pm1.3$, and the 
N$_2$D$^+$/N$_2$H$^+$ column density ratio was found to be $\sim0.6\pm0.1$, 
indicating a high degree of deuterium fractionation. 
In Paper III, we also presented the results of our SABOCA 350-$\mu$m imaging 
of dense cores in Orion B9. SMM 6, determined to be a thermally
supercritical cylindrical object, was resolved into three 
to four very low-mass (revised masses are $\sim0.2-0.6$ M$_{\odot}$; 
see Appendix~A) condensations projectively separated by $\sim0.06$ pc 
or $\sim1.2\times10^4$ AU (see Fig.~\ref{figure:SMM6})\footnote{We assumed that 
the distance to Orion B9 is $d=450$ pc. This value is also adopted in the 
present paper.}. The origin of this substructure can be explained in terms of 
thermal Jeans-type fragmentation. 
It seems likely that SMM 6 has undergone a similar 
fragmentation process as R CrA SMM 1A described above. 
In Table~\ref{table:sources}, we list the SABOCA 350-$\mu$m peak positions of 
the SMM 6's condensations along with their main physical properties. 

\begin{figure*}
\begin{center}
\includegraphics[scale=0.45]{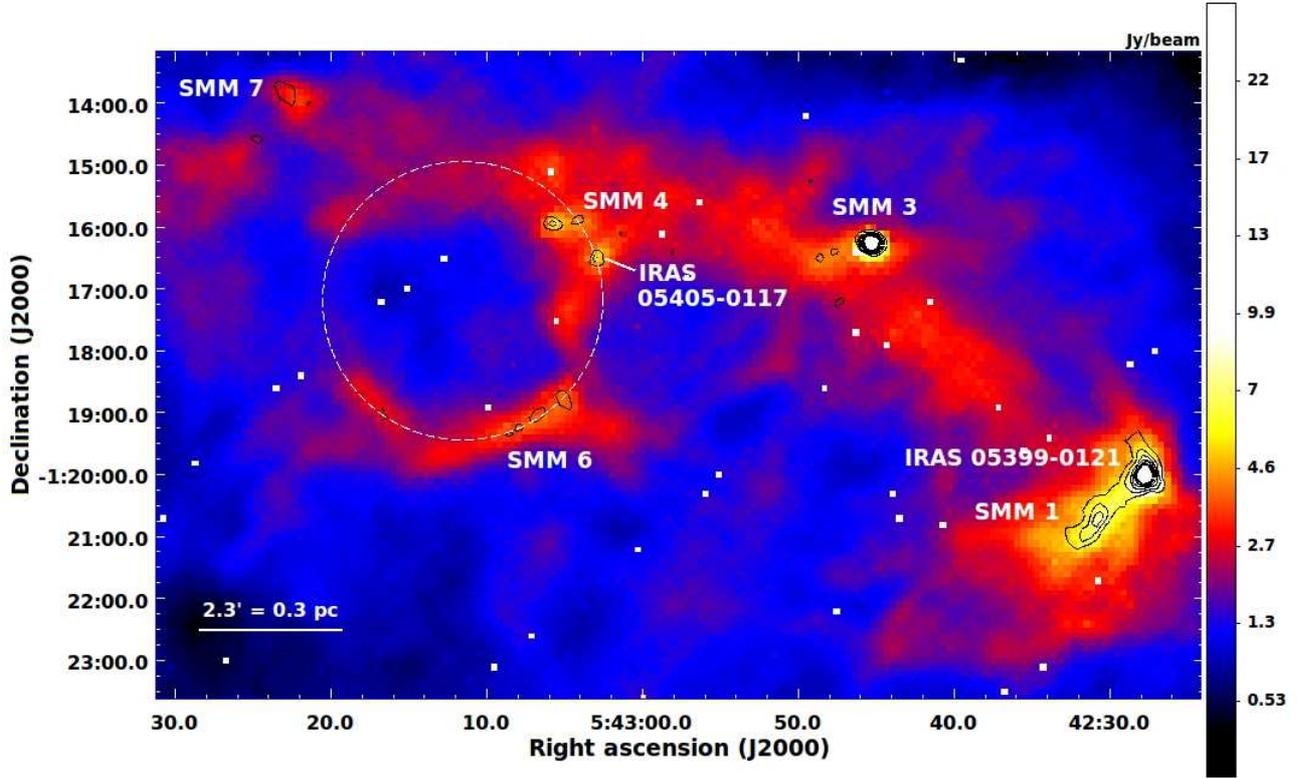}
\caption{\textit{Herschel}/SPIRE 250-$\mu$m far-infrared image towards the 
Orion B9 star-forming filament. The image is shown with a square-root scaling to
improve the contrast between bright and faint features. The image is overlaid 
with black contours of SABOCA 350-$\mu$m emission. The SABOCA contours start at
$3\sigma$ and increase at this interval (Paper III). The SABOCA data 
towards the IRAS 05399/SMM 1 -system with a $1\sigma$ noise level of 80 
mJy~beam$^{-1}$ are taken from Miettinen \& Offner (2013). 
Selected cores are labelled. The ring-like structure comprising partly of 
SMM 6 is highlighted with a white dashed circle of radius $2\farcm25$ or 
$\sim0.3$ pc. A scale bar indicating the 0.3 pc projected length is shown in 
the bottom left corner.}
\label{figure:spire}
\end{center}
\end{figure*}

Recent \textit{Herschel} data show that SMM 6 belongs to a northeast-southwest 
oriented filamentary structure (see Fig.~\ref{figure:spire}). 
Miettinen (2012b) found that there is a sharp velocity gradient in the parent 
filament (across its short axis), and suggested that it might 
represent a shock front resulting from the feedback from the nearby expanding 
\ion{H}{ii} region NGC 2024 ($\sim4$ pc southwest of Orion B9). The formation 
of dense cores in Orion B9, including SMM 6, might have been triggered by this 
feedback. As highlighted in Fig.~\ref{figure:spire}, the ring-like 
cloud structure consisting partly of SMM 6 could be a manifestation of such a 
dynamic environment.

In Paper III, we noticed that our LABOCA map employed in Papers I and II was 
misaligned, and therefore we had to adjust its pointing using the 
\textit{Spitzer} and SABOCA source positions (see footnote~2 in 
Paper III and \cite{miettinen2013}). The target positions 
of our previous molecular-line observations were chosen to be the peak 
positions of the LABOCA map before adjusting the pointing, and therefore they 
are slightly offset from the 870-$\mu$m maxima. In the case of SMM 6, our 
previous line observations probed the core edge (as shown by the green plus 
sign in Fig.~1). To better understand the physics and chemistry of SMM 6, 
particularly the properties of its substructure, we performed follow-up
molecular-line observations towards the condensations. 
In the present paper, we examine the kinematics and dynamics of these 
subfragments and their chemical properties, namely CO depletion and molecular 
deuteration of N$_2$H$^+$. The latter parameter has the potential to constrain 
the evolutionary stage of the condensations (e.g, \cite{crapsi2005}; 
\cite{friesen2013}). The present work is one of the first studies of 
the dynamics of subfragments within a prestellar core. The observations and 
data reduction are described in Sect.~2. Section~3 presents the observational 
results, while analysis and its results are presented in Sect.~4 and 
Appendix~A. We discuss the results in Sect.~5 and in Sect.~6, we summarise the 
paper and draw our main conclusions.

\begin{table*}
\caption{The 350-$\mu$m condensations within SMM 6.}
\begin{minipage}{2\columnwidth}
\centering
\renewcommand{\footnoterule}{}
\label{table:sources}
\begin{tabular}{c c c c c c c}
\hline\hline 
Source & $\alpha_{2000.0}$ & $\delta_{2000.0}$ & $R_{\rm eff}$ & $M$ & $N({\rm H_2})$ & $\langle n({\rm H_2}) \rangle$\\
       & [h:m:s] & [$\degr$:$\arcmin$:$\arcsec$] & [pc] & [M$_{\sun}$] & [$10^{22}$ cm$^{-2}$] & [$10^5$ cm$^{-3}$]\\
\hline
SMM 6a\tablefootmark{a} & 05 43 05.0 & -01 18 49.3 & 0.02 & $0.6\pm0.2$ & $ 2.7\pm1.3$ & $3.4\pm1.1$ \\
SMM 6b & 05 43 06.8 & -01 19 02.8 & 0.02 & $0.3\pm0.1$ & $2.2\pm0.8$ & $1.7\pm0.6$ \\
SMM 6c & 05 43 08.0 & -01 19 13.2 & 0.005\tablefootmark{b} & $0.1\pm0.05$\tablefootmark{b} & $1.7\pm0.7$ & $37\pm18$\tablefootmark{b} \\
SMM 6d & 05 43 08.6 & -01 19 22.3 & 0.005\tablefootmark{b} & $0.1\pm0.05$\tablefootmark{b} & $1.9\pm0.7$ & $37\pm18$\tablefootmark{b} \\
SMM 6c+6d\tablefootmark{b} & \ldots & \ldots & 0.01\tablefootmark{b} & $0.2\pm0.1$\tablefootmark{b} & \ldots & $9.1\pm4.6$\tablefootmark{b}\\
\hline 
\end{tabular} 
\tablefoot{Columns (2) and (3) give the equatorial coordinates 
[$(\alpha, \,\delta)_{2000.0}$] of the SABOCA peak position. Columns (4)--(7) 
list, respectively, the effective radius, mass, beam-averaged peak H$_2$ 
column density, and the volume-averaged H$_2$ number density. The latter three 
are revised from those presented in Paper III (see Appendix~A for 
details).\tablefoottext{a}{SMM 6a was called SMM 6 in Paper III. It is renamed 
here for clarity because we call the parent core 
SMM 6.}\tablefoottext{b}{With the {\tt clumpfind} 
(\cite{williams1994}) settings used in Paper III, SMM 6c and 6d were treated as 
a single source. The values of $R_{\rm eff}$, $M$, and 
$\langle n({\rm H_2}) \rangle$ given in the last row of the table 
include contributions from both condensations. Approximately half of the 
reported size and mass of SMM 6c+6d can be assigned to both 
fragments, making their densities about four times the common value 
($\sim3.7\pm1.8\times10^6$ cm$^{-3}$).}}
\end{minipage}
\end{table*}


\section{Observations and data reduction}

We used the APEX (Atacama Pathfinder EXperiment; \cite{gusten2006}) 12-m 
telescope in Chile to observe the spectral-line transitions C$^{17}$O$(2-1)$, 
N$_2$H$^+(3-2)$, and N$_2$D$^+(3-2)$ towards the 350-$\mu$m peaks of the dense 
condensations within SMM 6. The N$_2$H$^+$ observations were carried out on 25 
September 2012, while those of C$^{17}$O and N$_2$D$^+$ were performed on the 
27th of the month. The strongest 350-$\mu$m condensation SMM 6a was 
observed again in C$^{17}$O on 28 September because the previous day's 
weather was poor when the source was observed. 

The observed transitions, their selected spectroscopic pro\-perties, and 
observational parameters are summarised in Table~\ref{table:lines}. 
The critical densities, $n_{\rm crit}=A_{\rm ul}/C_{\rm ul}$, listed in Col.~(4) of 
Table~\ref{table:lines}, were calculated at $T=10$ K using the Einstein 
$A-$coefficients from the Leiden Atomic and Molecular Database (LAMDA; 
\cite{schoier2005})\footnote{{\tt http://www.strw.leidenuniv.nl/$\sim$moldata/}} (C$^{17}$O) and Pagani et al. (2009) (N$_2$H$^+$ and N$_2$D$^+$). The 
collisional-rate data ($C_{\rm ul}$) were taken from LAMDA (C$^{17}$O) and 
the Caltech Submillimeter Wave Astrophysics 
website\footnote{{\tt http://www.submm.caltech.edu/$\sim$tab/molecular$_{-}$data/}\\ {\tt collisional$_{-}$rates.html }} (N$_2$H$^+$ and N$_2$D$^+$). 
The latter are based on data from BASECOL 
(\cite{dubernet2013})\footnote{{\tt http://basecol.obspm.fr}} .

As a frontend for the C$^{17}$O$(2-1)$ and N$_2$D$^+(3-2)$ 
observations, we used the APEX-1 receiver of the Swedish Heterodyne Facility 
Instrument (SHeFI; \cite{belitsky2007}; \cite{vassilev2008a},b). 
APEX-1 operates in single-sideband (SSB) mode using sideband separation 
mixers, and it has a sideband rejection ratio $>10$ dB. 
For the N$_2$H$^+(3-2)$ observations the frontend 
used was APEX-2, which has similar cha\-racteristics as APEX-1. The backend for 
all observations was the RPG eXtended bandwidth Fast Fourier Transfrom 
Spectrometer (XFFTS; cf. \cite{klein2012}) with an instantaneous bandwidth of
2.5 GHz and 32\,768 spectral channels.

The observations were performed in the wobbler-switching mode with a 
$100\arcsec$ azimuthal throw (symmetric offsets) and a chopping rate of 0.5 Hz.
The telescope focus and pointing were optimised and checked at regular
intervals on the planet Jupiter, and the variable stars W Orionis, RAFGL865 
(V1259 Ori), $o$ Ceti (Mira A), and R Leporis (Hind's Crimson Star).
The pointing was found to be accurate to $\sim3\arcsec$. Calibration was made 
by means of the chopper-wheel technique and the output 
intensity scale given by the system is $T_{\rm A}^{\star}$, i.e., the 
antenna temperature corrected for the atmospheric attenuation. The observed 
intensities were converted to the main-beam brightness 
temperature scale by $T_{\rm MB}=T_{\rm A}^{\star}/\eta_{\rm MB}$, where 
$\eta_{\rm MB}$ is the main-beam efficiency. The absolute calibration 
uncertainty is estimated to be about 10\%.

The spectra were reduced using the CLASS90 programme of the GILDAS software 
package\footnote{Grenoble Image and Line Data Analysis Software is provided 
and actively developed by IRAM, and is available at 
{\tt http://www.iram.fr/IRAMFR/GILDAS}}. 
The individual spectra were averaged and the resulting spectra were 
Hanning-smoothed to improve the signal-to-noise ratio. 
Linear (first-order) baselines were determined from
velocity ranges without line-emission features, and then subtracted
from the spectra. The resulting $1\sigma$ rms noise levels at the smoothed 
resolutions are listed in the last column of Table~\ref{table:lines}.  

The $J=2-1$ transition of C$^{17}$O contains nine hyperfine (hf) components 
that are spread over $\sim2.4$ km~s$^{-1}$. We fitted this hf structure 
using ``method hfs'' of CLASS90 to derive the LSR velocity 
(${\rm v}_{\rm LSR}$) of the emission, and FWHM linewidth ($\Delta {\rm v}$). 
The hf line fitting can also be used to derive the line 
optical thickness by comparing the relative intensities of the hf 
components. However, in all spectra the hf components are 
mostly blended together and thus the total optical thickness could 
not be reliably determined from the sum of the peak optical 
thicknesses of the components. For the rest frequencies of the hf components, 
we used the values from Ladd et al. (1998; Table 6 therein). The adopted 
central frequency, 224\,714.199 MHz, refers to the strongest hf component 
$J_F = 2_{9/2} \rightarrow 1_{7/2}$ which has a relative intensity of 
$R_i=\frac{1}{3}$.

The $J=3-2$ transition of both N$_2$H$^+$ and N$_2$D$^+$ is split up into 38 
hf components spread over about 5.7 and 6.9 km~s$^{-1}$, respectively. 
To fit these hf structures, we used the rest frequencies from
Pagani et al. (2009; Tables 4 and 10 therein). The adopted central 
frequencies of N$_2$H$^+(3-2)$ and N$_2$D$^+(3-2)$, 279\,511.832 and 
231\,321.912 MHz, are those of the $J_{F_1 F} = 3_{45} \rightarrow 2_{34}$
hf component which has a relative intensity of $R_i=\frac{11}{63}$.
Also in these cases, the hf components are blended and thus the optical 
thickness could not be reliably determined through hf fitting.

\begin{table*}
\caption{Observed spectral-line transitions and observational parameters.}
\begin{minipage}{2\columnwidth}
\centering
\renewcommand{\footnoterule}{}
\label{table:lines}
\begin{tabular}{c c c c c c c c c c c c}
\hline\hline 
Transition & $\nu$ & $E_{\rm u}/k_{\rm B}$ & $n_{\rm crit}$ & HPBW & $\eta_{\rm MB}$ & $T_{\rm sys}$ & PWV & \multicolumn{2}{c}{Channel spacing\tablefootmark{a}} & $\tau_{\rm int}$ & rms\\
      & [MHz] & [K] & [cm$^{-3}$] & [\arcsec] & & [K] & [mm] & [kHz] & [km~s$^{-1}$] & [min] & [mK]\\
\hline        
C$^{17}$O$(2-1)$ & 224\,714.199\tablefootmark{b} & 16.2 & $2.1\times10^4$ & 27.8 & 0.75 & 168--208 & 1.0--2.1 & 76.3 & 0.1 & 9 & 27--34\\
N$_2$D$^+(3-2)$ & 231\,321.912\tablefootmark{c} & 22.2 & $3.0\times10^6$& 27.0 & 0.75 & 204--228 & 1.0--2.1 & 76.3 & 0.1 & 9--15 & 30--36\\ 
N$_2$H$^+(3-2)$ & 279\,511.832\tablefootmark{c} & 26.8 & $5.2\times10^6$ & 22.3 & 0.74 & 161--164 & $\sim0.2$ & 76.3 & 0.08 & 4.2--4.8 & 38--41\\
\hline 
\end{tabular} 
\tablefoot{Columns (2)--(4) give the rest frequencies of the observed 
transitions ($\nu$), their upper-state energies ($E_{\rm u}/k_{\rm B}$, 
where $k_{\rm B}$ is the Boltzmann constant), and critical densities (see text 
for details). Columns (5)--(12) give the APEX beamsize (HPBW) and the 
main-beam efficiency ($\eta_{\rm MB}$) at the observed frequencies, the SSB 
system temperatures during the observations ($T_{\rm sys}$ on a $T_{\rm MB}$ 
scale, see text), the amount of PWV, channel widths 
(both in kHz and km~s$^{-1}$) of the original data, the on-source integration 
times per position ($\tau_{\rm int}$), and the $1\sigma$ rms noise levels at 
the smoothed resolution.\\
\tablefoottext{a}{The original channel spacings. The final spectra were 
Hanning-smoothed which divides the number of channels by two.}\tablefoottext{b}{From Ladd et al. (1998).} \tablefoottext{c}{From Pagani et al. (2009).}}
\end{minipage}
\end{table*}

\section{Observational results}

\subsection{Spectra}

\begin{figure*}
\begin{center}
\includegraphics[width=0.247\textwidth]{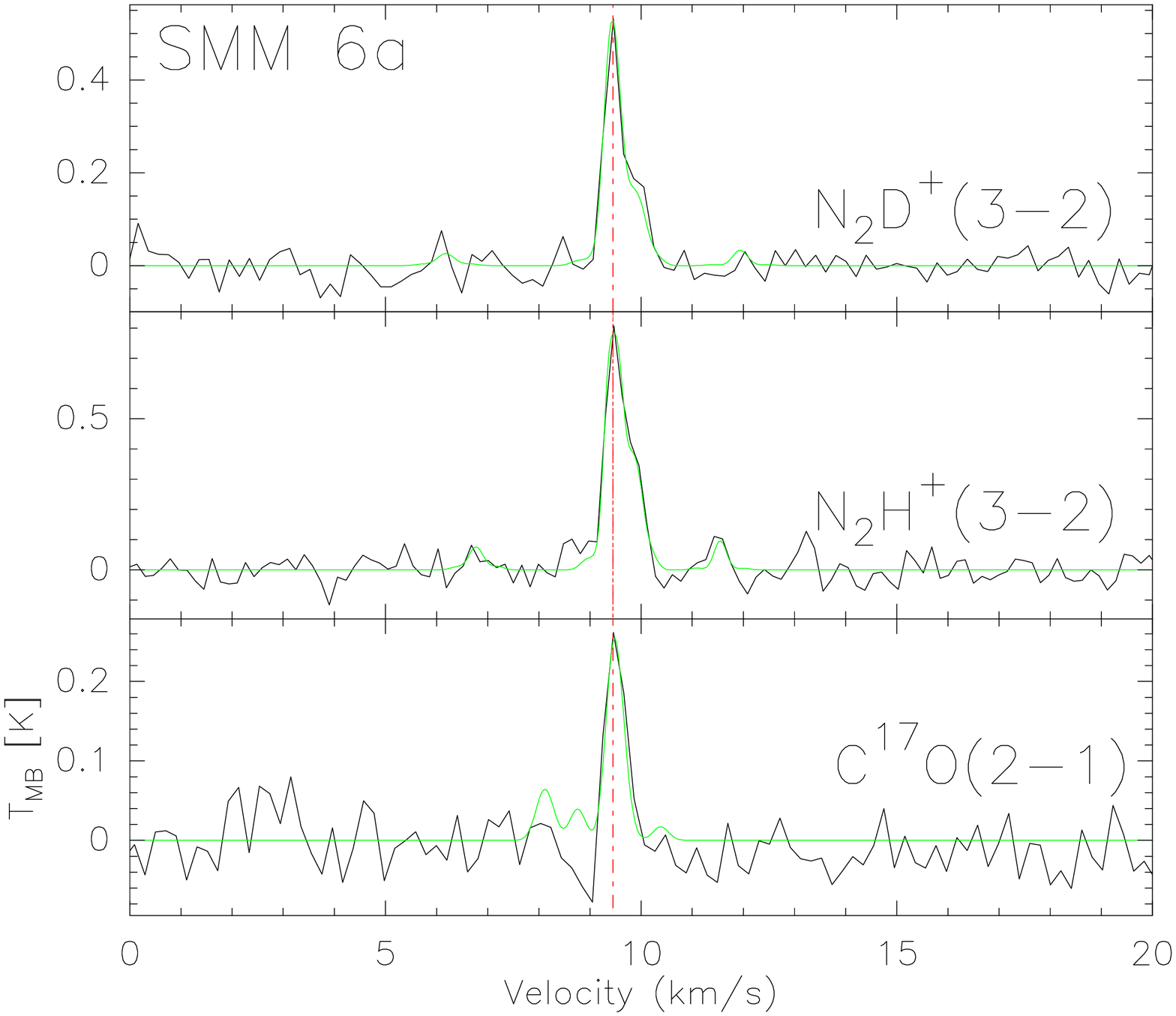}
\includegraphics[width=0.247\textwidth]{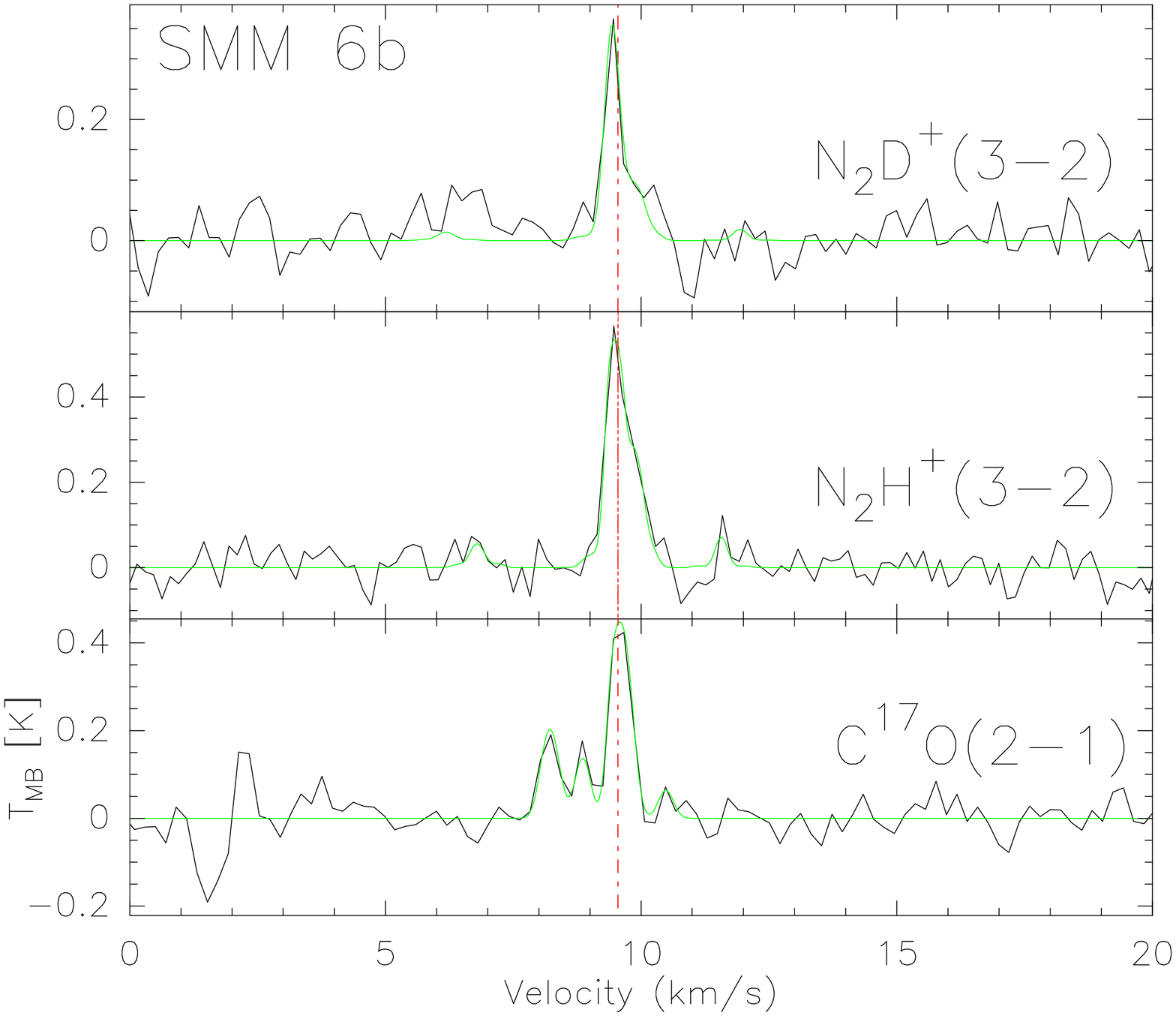}
\includegraphics[width=0.247\textwidth]{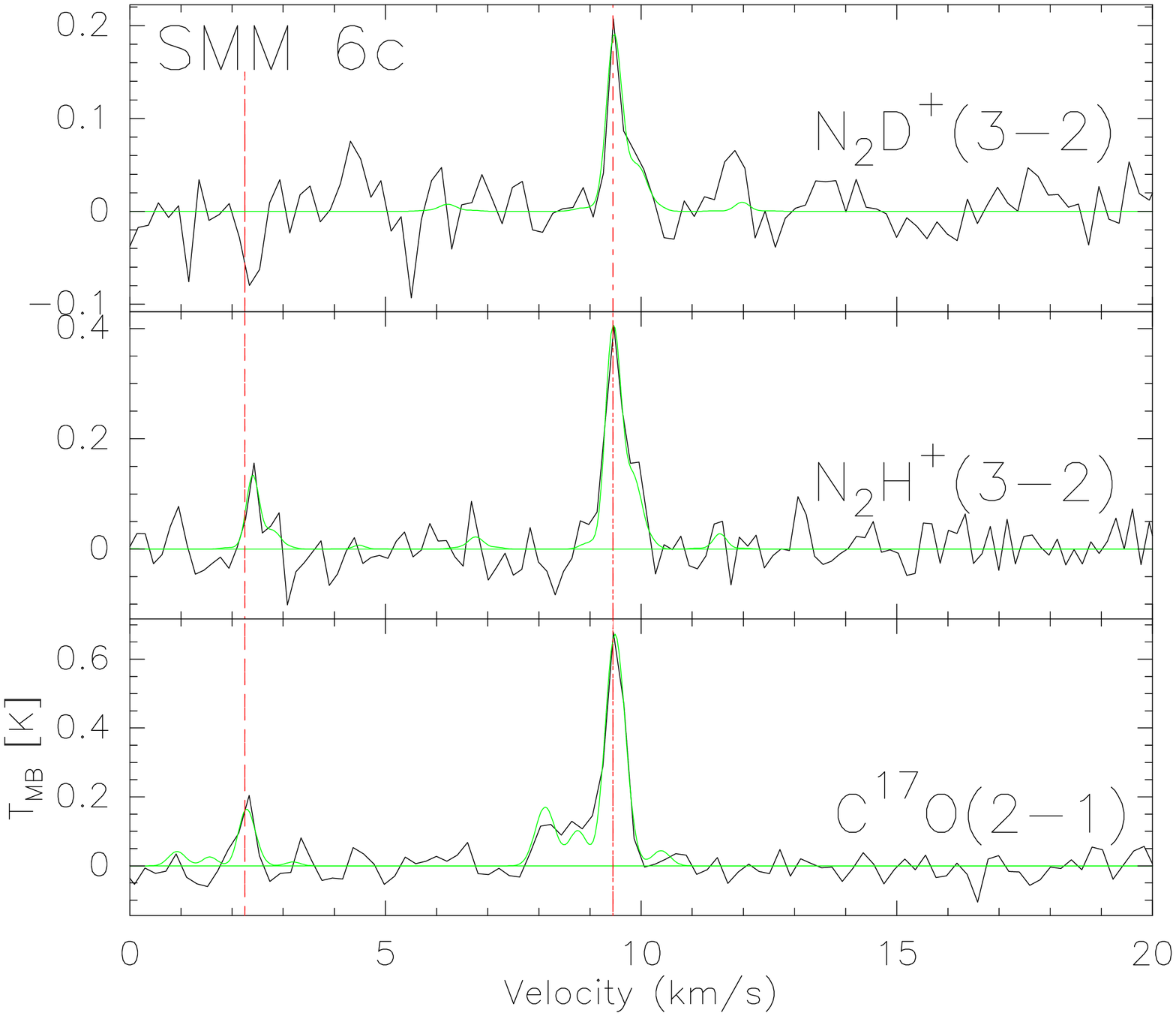}
\includegraphics[width=0.247\textwidth]{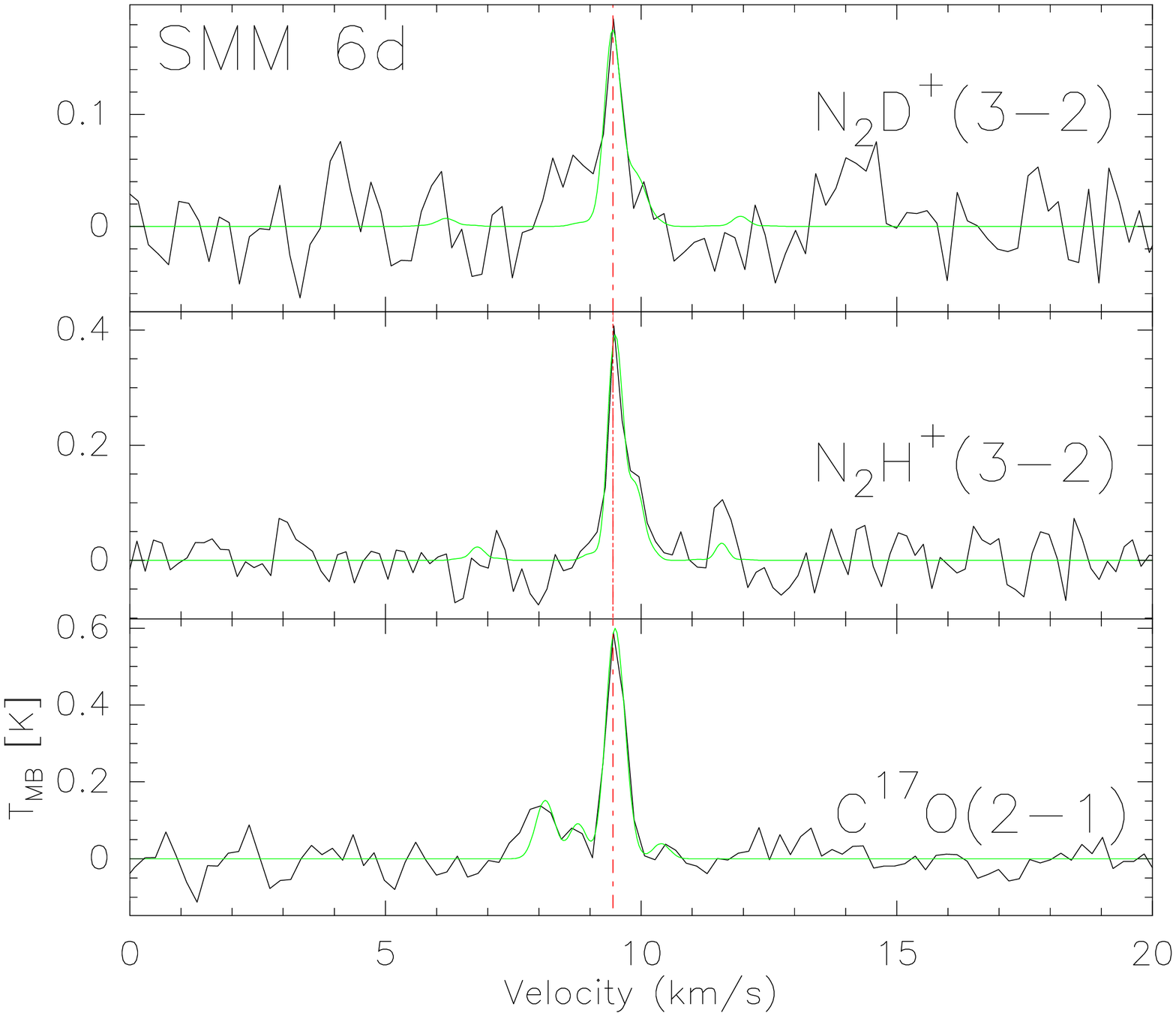}
\caption{Hanning-smoothed C$^{17}$O$(2-1)$, N$_2$H$^+(3-2)$, and 
N$_2$D$^+(3-2)$ spectra. Hyperfine-structure fits to the lines are overlaid in 
green. In each panel, the vertical red dashed line indicates the radial 
velocity of the C$^{17}$O$(2-1)$ line. Note that towards SMM 6c, two velocity 
components  of C$^{17}$O$(2-1)$ and N$_2$H$^+(3-2)$ (at $\sim2.3$ and $\sim9.4$ 
km~s$^{-1}$) are detected.}
\label{figure:spectra}
\end{center}
\end{figure*}

The Hanning-smoothed spectra are presented in Fig.~\ref{figure:spectra}. 
The lines are clearly detected towards all condensations. The hf structure of
the C$^{17}$O$(2-1)$ line is partially resolved in SMM 6b--d (note the 
hf satellites on the low-velocity side of the strongest component), but not 
sufficiently well to derive the line optical thickness. Towards SMM 6c, 
the C$^{17}$O and N$_2$H$^+$ lines exhibit two well-separated velocity 
components: besides the lines at the systemic velocity $\sim9.4$ km~s$^{-1}$, 
there are weak lines at a lower velo\-city ($\sim2.3$ km~s$^{-1}$). 
There is also a hint of low-velocity C$^{17}$O emission in the spectra towards 
SMM 6a and 6b. The low-velocity line emission was already detected towards 
other Orion B9 cores in Papers I--III and by Miettinen (2012b). 
We note that some of the spectra show small artificial ``absorption''-like 
features, most notably the C$^{17}$O spectra towards SMM 6a and 6b. These are 
caused by emission in the OFF-source reference position (OFF-beam); 
for example, there is C$^{17}$O emission in the velocity range $\sim1-2$ 
km~s$^{-1}$ towards the OFF-position of SMM 6b. We also note that the nearby 
condensations SMM 6c and 6d are encompassed by the $22\farcs3-27\farcs8$ 
beams. Therefore, line emission from one of the sources is (partly) caught in 
the beam when observing towards the dust peak of the other and may thus 
affect the line profiles to some degree.   

\subsection{Spectral-line parameters}

The spectral-line parameters are listed in Table~\ref{table:lineparameters}.
The values of ${\rm v}_{\rm LSR}$ and $\Delta {\rm v}$ were derived through 
fitting the hf structure. The quoted uncertainties are formal $1\sigma$ 
fitting errors. The fact that the lines are only marginally
resolved at best means that the quoted $\Delta {\rm v}$ values represent only 
an upper limit to the intrinsic linewidth. 
The peak intensities ($T_{\rm MB}$) were derived by fitting a 
single Gaussian to the profile of the strongest line. 
The integrated line intensities ($\int T_{\rm MB} {\rm dv}$) listed 
in Col.~(6) of Table~\ref{table:lineparameters} were computed over the 
velocity range given in square brackets in the corresponding column. 
This way, we were able to take the non-Gaussian shape of the lines 
into account. The uncertainties in the peak and integrated intensities 
represent the quadrature sums of the fitting errors and the 10\% calibration 
uncertainties. In the last two columns of Table~\ref{table:lineparameters}, 
we list the estimated peak line optical-thicknesses ($\tau_0$) and assumed 
excitation temperatures ($T_{\rm ex}$), which are described in Sect~4.5.

\begin{table*}
\caption{Spectral-line parameters.}
{\small
\begin{minipage}{2\columnwidth}
\centering
\renewcommand{\footnoterule}{}
\label{table:lineparameters}
\begin{tabular}{c c c c c c c c}
\hline\hline 
Source & Transition & ${\rm v}_{\rm LSR}$ & $\Delta {\rm v}$ & $T_{\rm MB}$ & $\int T_{\rm MB} {\rm dv}$\tablefootmark{a} & $\tau_0$\tablefootmark{b} & $T_{\rm ex}$\tablefootmark{c}\\
     & & [km~s$^{-1}$] & [km~s$^{-1}$] & [K] & [K~km~s$^{-1}$] & & [K]\\
\hline
SMM 6a & C$^{17}$O$(2-1)$ & $9.45\pm0.02$ & $0.36\pm0.05$ & $0.28\pm0.05$ & $0.12\pm0.02$ [9.09, 10.46; 73.4\%] & $0.05\pm0.01$ & 10.0 \\
      & N$_2$H$^+(3-2)$ & $9.42\pm0.01$ & $0.27\pm0.02$ & $0.71\pm0.11$ & $0.56\pm0.06$ [6.26, 12.00; 99.9\%] & $1.62\pm0.28$ & 5.0\\
      & N$_2$D$^+(3-2)$ & $9.38\pm0.01$ & $0.33\pm0.01$ & $0.47\pm0.10$ & $0.30\pm0.04$ [8.22, 11.09; 92.6\%] & $0.52\pm0.04$ & 5.0\\
SMM 6b & C$^{17}$O$(2-1)$ & $9.55\pm0.02$ & $0.34\pm0.02$ & $0.48\pm0.06$ & $0.39\pm0.05$ [7.69, 10.74; 100\%] & $0.09\pm0.01$ & 10.0\\
       & N$_2$H$^+(3-2)$ & $9.44\pm0.01$ & $0.27\pm0.03$ & $0.49\pm0.08$ & $0.37\pm0.04$ [7.90, 10.67; 92.6\%] & $0.81\pm0.06$ & 5.0\\
       & N$_2$D$^+(3-2)$ & $9.38\pm0.02$ & $0.33\pm0.04$ & $0.35\pm0.05$ & $0.21\pm0.03$ [8.46, 10.67; 92.6\%] & $0.36\pm0.01$ & 5.0\\
SMM 6c & C$^{17}$O$(2-1)$ & $9.45\pm0.01$ & $0.42\pm0.03$ & $0.68\pm0.08$ & $0.46\pm0.05$ [7.66, 10.42; 100\%] & $0.14\pm0.01$ & 10.0\\
SMM 6c (2nd v-comp.) & C$^{17}$O$(2-1)$ & $2.25\pm0.10$ & $0.34\pm0.29$ & $0.21\pm0.02$ & $0.06\pm0.04$ [1.50, 2.79; 78.7\%] & $0.04\pm0.01$ & 10.0\\
            & N$_2$H$^+(3-2)$ & $9.41\pm0.02$ & $0.27\pm0.04$ & $0.34\pm0.06$ & $0.23\pm0.03$ [8.57, 11.72; 96.1\%] & $0.48\pm0.02$ & 5.0 \\
SMM 6c (2nd v-comp.) & N$_2$H$^+(3-2)$ & $2.36\pm0.03$ & $0.27\pm0.03$ & $0.16\pm0.05$ & $0.05\pm0.01$ [1.92, 3.04; 89.9\%] & $0.20\pm0.01$ & 5.0 \\
       & N$_2$D$^+(3-2)$ & $9.42\pm0.03$ & $0.33\pm0.02$ & $0.21\pm0.02$ & $0.10\pm0.02$ [8.53, 10.32; 90.9\%] & $0.20\pm0.01$ & 5.0\\
SMM 6d & C$^{17}$O$(2-1)$ & $9.45\pm0.01$ & $0.39\pm0.03$ & $0.60\pm0.06$ & $0.42\pm0.05$ [6.82, 10.98; 100\%] & $0.12\pm0.01$ & 10.0\\
       & N$_2$H$^+(3-2)$ & $9.45\pm0.01$ & $0.27\pm0.02$ & $0.32\pm0.07$ & $0.26\pm0.03$ [8.22, 12.03; 96.2\%] & $0.45\pm0.03$ & 5.0\\
       & N$_2$D$^+(3-2)$ & $9.38\pm0.04$ & $0.35\pm0.11$ & $0.17\pm0.02$ & $0.14\pm0.04$ [7.73, 10.53; 87.2\%] & $0.16\pm0.01$ & 5.0\\
\hline 
\end{tabular} 
\tablefoot{
\tablefoottext{a}{Integrated intensity was computed by integrating over the 
velocity range indicated in square brackets (in km~s$^{-1}$). The percentage
value indicates the contribution of hf components' intensity lying within the 
corresponding velocity range.}\tablefoottext{b}{The optical thickness in the 
centre of the strongest hf component; it was derived using the assumed 
$T_{\rm ex}$ value given in the last column.}\tablefoottext{c}{$T_{\rm ex}$ is 
assumed to be the same for the $J=3-2$ transition of N$_2$H$^+$ and N$_2$D$^+$.
See Sect.~4.5 for details.}}
\end{minipage} }
\end{table*}

\section{Analysis and results}

\subsection{Revisiting the core fragmentation}

Because the mass of the parent core is revised in the present work 
(Appendix~A), we also re-analyse its fragmentation here. 


The line mass of SMM 6 is found to be 
$M_{\rm line}\simeq 27\pm6$ M$_{\sun}$~pc$^{-1}$. For an infinite, 
unmagnetised, isothermal cylinder, instability (collapse to line singularity) 
occurs if its $M_{\rm line}$ exceeds the critical equilibrium value of 
(e.g., \cite{ostriker1964}; \cite{inutsuka1992})

\begin{equation}
\label{eq:linemass}
M_{\rm line}^{\rm crit}=\frac{2c_{\rm s}^2}{G}\,,
\end{equation}
where $c_{\rm s}$ is the isothermal sound speed and $G$ the gravitational 
constant. Using the gas temperature $T_{\rm kin}=11.0\pm0.4$ K to calculate the 
sound speed ($0.196\pm0.004$ km~s$^{-1}$), we derive the value 
$M_{\rm line}^{\rm crit}=17.8\pm0.6$ M$_{\sun}$~pc$^{-1}$. If the 
$\sim0.1$ km~s$^{-1}$ non-thermal velocity dispersion derived in Paper 
II is taken into account, the effective sound speed 
$c_{\rm eff}=(c_{\rm s}^2+\sigma_{\rm NT}^2)^{1/2}$ increases the value of 
$M_{\rm line}^{\rm crit}$ to $\sim 22.5\pm0.8$ M$_{\sun}$~pc$^{-1}$. 
SMM 6 therefore appears to be (slightly) supercritical, in accordance 
with the detected substructure along the filament.

The projected spacing between the condensations is $\sim29\arcsec$ or
$\sim0.06$ pc, where the sources 6c and 6d are treated as a single
fragment. In the case the cylinder has $M_{\rm line}=M_{\rm line}^{\rm crit}$ as is 
roughly the case for SMM 6, the Jeans length along the long axis is 
(\cite{larson1985}; \cite{hartmann2002})

\begin{equation}
\lambda_{\rm J}^{\rm cyl}=\frac{3.94c_{\rm s}^2}{G\Sigma_0}\,,
\end{equation}
where $\Sigma_0=\mu_{\rm H_2}m_{\rm H}N({\rm H_2})$ is the central 
surface density with $\mu_{\rm H_2}=2.82$ being the mean molecular weight per 
H$_2$ molecule (when the He/H abundance ratio is about 0.1; see Appendix~A) 
and $m_{\rm H}$ the mass of a hydrogen atom. 
If the central column density is adopted to be the average value 
derived towards the condensations, $\sim 2.1\pm0.2\times10^{22}$ cm$^{-2}$ 
[Col.~(6) of Table~\ref{table:sources}], $\lambda_{\rm J}^{\rm cyl}$ is 
$\simeq0.07\pm0.01$ pc -- very close to the observed interval of 
the condensations. We note that due to projection effects we are 
measuring only a lower limit to the separations. 

If the \textit{finite} length and diameter of the cylindrical cloud are $L$ 
and $D$, respectively, the number of fragments forming due to gravitational 
instability is expected to be (e.g., \cite{bastien1991}; \cite{wiseman1998})

\begin{equation}
N_{\rm frag}=\frac{L}{\lambda_{\rm crit}}=\frac{2\left(L/D \right)}{3.94}\simeq0.5A\,,
\end{equation}
where $\lambda_{\rm crit}=1.97D$ is the wavelength of the most unstable 
perturbation, and $A$ the cylinder's aspect ratio. The average 
width of the SMM 6 filament is about $26\farcs4$ or $D=2R=0.06$ pc, so $A$ 
is about 4. One would therefore expect to have two subfragments, i.e., 
half the number we observe. The discrepancy could be caused by 
projection effects (the true length of the filament is longer), and the fact 
that the source is not a perfect cylinder but is bent or curved as shown in 
Fig.~\ref{figure:spire} and it narrows towards the southeast (i.e., $D$ is
changing along the filament). The number of detected subfragments could also be
an imprint of internal turbulent motions, i.e., turbulent fragmentation 
(cf.~\cite{nakamura2012}).
Note that $\lambda_{\rm crit}$ is about 0.12 pc when computed using the average 
width 0.06 pc, i.e., a factor of 1.7 larger than $\lambda_{\rm J}^{\rm cyl}$. 

We conclude that SMM 6 has likely fragmented into subcondensations as 
a result of cylindrical Jeans-type gravitational instability. 
Considering a filament of radius $R$, the fragmentation timescale is then 
expected to be comparable to the radial crossing time, 
$\tau_{\rm cross}=R/\sigma$, where $\sigma$ refers to 
the total velocity dispersion. If we use the average FWHM linewidth of the 
C$^{17}$O lines detected here, $\Delta {\rm v}=\sqrt{8 \ln 2}\sigma=0.38$ 
km~s$^{-1}$, the $\tau_{\rm cross}$ for the SMM 6 filament is 
$\sim1.8\times10^5$ yr.

\subsection{Gas velocity dispersion and virial state of the condensations}

To study the gas kinematics of the high-density portion of the condensations, 
we employed the N$_2$H$^+(3-2)$ transition due to its high critical density. 
We computed the non-thermal portion of the line-of-sight velocity dispersion, 
$\sigma_{\rm NT}$, using Eq.~(9) of Paper II. It was assumed that 
$T_{\rm kin}=10$ K, and the error in $\sigma_{\rm NT}$ was calculated from that of 
the FWHM linewidth. As shown in Col.~(2) of Table~\ref{table:virial}, 
$\sigma_{\rm NT}=0.1$ km~s$^{-1}$ for each source. Because the isothermal sound 
speed at 10 K is $c_{\rm s}=0.19$ km~s$^{-1}$, the 
$\sigma_{\rm NT}/c_{\rm s}$ ratios are $\sim0.5\pm0.1$, i.e., the condensations 
are characterised by \textit{subsonic} internal non-thermal (turbulent) motions.
The true values of $\sigma_{\rm NT}$ could be even lower because the 
N$_2$H$^+(3-2)$ lines are not spectrally resolved, enabling us to derive only 
upper limits to the linewidths and velocity dispersions.

To examine the dynamical state of the condensations, we calculated
their virial masses by employing the N$_2$H$^+(3-2)$ linewidths and 
using the formula where the effects of external pressure, rotation, 
and magnetic field are ignored [cf.~Eq.~(12) of Paper II]:

\begin{equation}
M_{\rm vir}=\frac{5}{8\ln 2}\frac{R_{\rm eff}\Delta {\rm v}^2}{aG}\, ,
\end{equation}
where the dimensionless parameter $a$ measures the effect of a
non-uniform density distribution on the gravitational energy and is given by 
$a=(1 - p/3)/(1 - 2p/5)$, where $p$ is the power-law index of the density 
profile ($n \propto r^{-p}$; \cite{bertoldi1992}). Following ABMP07, we 
assume that the density structure of the sources is that of a centrally 
condensed isothermal sphere with $n(r) \propto r^{-2}$ ($a=5/3$). 
The resulting values are listed in Col.~(3) of 
Table~\ref{table:virial} with uncertainties propagated from the linewidth 
errors. We note that if the condensations have inner density profiles 
shallower than $\sim r^{-2}$, the virial masses will be higher. For example, 
if the density power-law index is $p=1$ or $a=10/9$, the 
$M_{\rm vir}$ values are larger by a factor of 1.5 compared to those with $p=2$.

The virial parameters of the condensations were calculated following the 
definition by Bertoldi \& McKee (1992), i.e., $\alpha_{\rm vir}=M_{\rm vir}/M$. 
The values are given in Col.~(4) of Table~\ref{table:virial} with the 
uncertainties derived by propagating the errors in both mass estimates. 
The uncertainties in $\alpha_{\rm vir}$ are quite large, but gravitational 
boundedness with $\alpha_{\rm vir}<2$ seems possible at least for SMM 6a and 
6c,d. If the intrinsic N$_2$H$^+(3-2)$ linewidths are smaller, the 
$\alpha_{\rm vir}$ values would also become smaller enhancing the importance 
of the condensations' self-gravity.

\begin{table}
\caption{The one-dimensional non-thermal velocity dispersions, virial masses, 
and virial parameters of the condensations.}
\begin{minipage}{1\columnwidth}
\centering
\renewcommand{\footnoterule}{}
\label{table:virial}
\begin{tabular}{c c c c}
\hline\hline 
Source & $\sigma_{\rm NT}$ & $M_{\rm vir}$\tablefootmark{a} & $\alpha_{\rm vir}$ \\
        & [km~s$^{-1}$] & [M$_{\sun}$] &\\
\hline
SMM 6a & $0.10\pm0.01$ & $0.6\pm0.03$ & $1.0\pm0.4$ \\
SMM 6b & $0.10\pm0.01$ & $0.6\pm0.04$ & $2.1\pm0.7$ \\
SMM 6c & $0.10\pm0.02$ & $0.16\pm0.01$\tablefootmark{b} & $1.6\pm0.8$\tablefootmark{b} \\
SMM 6d & $0.10\pm0.01$ & $0.16\pm0.01$\tablefootmark{b} & $1.6\pm0.8$\tablefootmark{b} \\
SMM 6c+6d & \ldots & $0.3\pm0.02$\tablefootmark{c} & $1.6\pm0.8$\tablefootmark{c}\\
\hline 
\end{tabular} 
\tablefoot{The listed values assume that $T_{\rm kin}=10$ K.\tablefoottext{a}{A 
density profile that declines like $\sim r^{-2}$ was assumed.}\tablefoottext{b}{
These values assume that the radii and masses of SMM 6c and 6d are 
$\sim0.005$ pc and $\sim0.1$ M$_{\sun}$ (cf.~Table~\ref{table:sources}).}\tablefoottext{c}{For these values SMM 6c and 6d are treated as a one source.}}
\end{minipage} 
\end{table}

\subsection{Relative motions and interactions between the 
condensations}

The centroid velocity difference between C$^{17}$O and N$_2$H$^+$ for 
each condensation is small, only 0.045 km~s$^{-1}$ on average. 
This suggests that the condensations are not moving much with respect to the 
lower-density envelope (e.g., \cite{kirk2007}, 2010). The observed radial 
velocity data can also be used to study the (possible) motions of the 
condensations relative to one another (e.g., ABMP07). 
The average line-of-sight or radial velocity of the N$_2$H$^+(3-2)$ lines is 
9.43 km~s$^{-1}$ and the dispersion of the velocity centroids is 
$\sigma_{\rm v_{LSR},\,1D}^{\rm N_2H^+}\simeq0.02$ km~s$^{-1}$ 
($\langle {\rm v}_{\rm LSR}\rangle=9.48$ km~s$^{-1}$ and 
$\sigma_{\rm v_{LSR},\,1D}^{\rm C^{17}O}\simeq0.05$ km~s$^{-1}$ for the optically 
thin C$^{17}$O lines). The latter value is taken to represent the 
one-dimensional dispersion of the velocities of the high-density 
condensations (\cite{peng1998}; ABMP07). If the condensation 
velocities perpendicular to the line of sight are the same as their radial 
velocities, the three-dimensional dispersion of the velocities would be 
$\sigma_{\rm 3D}^{\rm N_2H^+}=\sqrt{3}\times\sigma_{\rm v_{LSR},\,1D}^{\rm N_2H^+}\simeq 0.03$ km~s$^{-1}$. If the orientation of the SMM 6 filament is 
\textit{not} exactly perpendicular to our line of sight, and \textit{if} the 
condensations would be moving along the filament, one would expect to see 
an imprint of these motions in the radial velocity data. The 
${\rm v}_{\rm LSR}$ values would be expected to change monotonically if the 
movement of condensations along the parent filament (in either direction) is 
taking place. However, such a systematic trend is not observed.


As mentioned in Appendix~A the length of SMM 6 along its long 
axis is $L=0.22$ pc. Using the above value of $\sigma_{\rm 3D}^{\rm N_2H^+}$, 
a typical timescale for the condensations to cross the long axis of the parent 
core becomes 
$\tau_{\rm cross}^{\rm long}=L/\sigma_{\rm 3D}^{\rm N_2H^+}\simeq7.2\times10^6$ yr. 
On the other hand, the average width of the filament, $D=2R=0.06$ pc, suggests 
a radial crossing time of 
$\tau_{\rm cross}^{\rm radial}=D/\sigma_{\rm 3D}^{\rm N_2H^+}\simeq2\times10^6$ yr. 
Because the projected separation between the condensations is also 
$\simeq0.06$ pc (when SMM 6c and 6d are treated as a one fragment), the time 
required for one condensation to encounter and interact with another 
condensation is expected to be $\sim2$ Myr (e.g., \cite{takakuwa2003}). 
We note that there are two effects that can affect the rate of 
collision between condensations: \textit{i)} gas drag caused by the 
surrounding medium, and \textit{ii)} gravitational focusing (ABMP07). 
While the former is likely to make the real collision timescale longer, the 
latter could enhance the collision rate. Moreover, gas gradients (density, 
velocity) in the parent filament (cf.~\cite{kirk2013}) could modify the time 
required for the condensations to encounter and interact with one another. 

If the representative density of the condensations is $\sim$a few 
$\times10^5$ cm$^{-3}$, their free-fall time is only 
$\tau_{\rm ff}\sim 7\times10^4$ yr. Therefore, it seems unlikely 
that the condensations have time to interact with one another before collapsing 
into protostars or proto-brown dwarfs unless their lifetime is a few tens of 
$\tau_{\rm ff}$, which seems unlikely (see, e.g., ABMP07, and 
references therein). Within the errors, it is in principle possible that the 
condensations are not gravitationally bound. For unbound condensations, 
one should also consider the dissipation timescale, 
$\tau_{\rm diss}=2R_{\rm eff}/c_{\rm eff}$ (e.g., \cite{takakuwa2003}). 
For the average condensation effective radius, 0.017 pc, and average 
$c_{\rm eff}$ calculated from the N$_2$H$^+(3-2)$ lines, 
0.215 km~s$^{-1}$, $\tau_{\rm diss}$ becomes $\sim1.5\times10^5$ yr. The typical 
dissipation speed of the condensations would therefore be an order of 
magnitude shorter than the coalescence timescale.

\subsection{Mass accretion onto the condensations}

Some numerical simulations suggest that protostars can gain mass via 
competitive accretion, i.e., accretion from a shared reservoir of material 
(\cite{bonnell2001}; \cite{bonnell2006}). Even though competitive accretion 
models apply to protostars that accrete the available gas while moving around 
within the protocluster potential well, we can still obtain a rough estimate 
of the accretion rate onto the starless condensations if competitive-like 
accretion is occurring (cf.~ABMP07).
The mass accretion rate is 
given by $\dot{M}_{\rm acc}\approx \pi \rho_0 V_{\rm rel} R_{\rm acc}^2$, where 
$\rho_0$ is the gas density of the background medium, $V_{\rm rel}$ is the 
relative velocity of the subfragments with respect to the ambient 
medium, and $R_{\rm acc}$ is the accretion radius [see Eq.~(3) of ABMP07].
The revised volume-average density of SMM 6 is about $\sim1.6\times10^4$ 
cm$^{-3}$, the value we adopt for the background gas density. 
We assume that $V_{\rm rel}$ is equal to 
$\sigma_{\rm 3D}^{\rm N_2H^+}\simeq 0.03$ km~s$^{-1}$.\footnote{ABMP07 
assumed that the velocity distribution of their sample of condensations 
was given by the Maxwell-Boltzmann distribution, and adopted 
$V_{\rm rel}$ for the average relative speed of a condensation 
(${\rm v}_{\rm ave}=\sqrt{8/\pi}\times \sigma_{\rm v_{LSR},\,1D}^{\rm N_2H^+}$ 
in our notation). The Maxwell-Boltzmann distribution is not expected 
to apply to our small number of subfragments, but we note that the value of 
$\sigma_{\rm 3D}^{\rm N_2H^+}$ is very close to ${\rm v}_{\rm ave}$.}
As concluded by ABMP07, $R_{\rm acc}$ is typically comparable to the 
condensation radius, and therefore we estimate it to be the average effective 
radius of the condensations, i.e., 0.017 pc. We thus derive the value 
$\dot{M}_{\rm acc}\sim2.6\times10^{-8}$ M$_{\sun}$~yr$^{-1}$. For comparison, 
the characteristic mass infall rate for a singular isothermal sphere undergoing 
gravitational collapse, $\sim c_{\rm s}^3/G$ (\cite{shu1977}), 
is $\sim1.5\times10^{-6}$ M$_{\sun}$~yr$^{-1}$ at 10 K -- about 60 times higher 
than the above $\dot{M}_{\rm acc}$ value. Local gravitational instability of the 
individual condensations is therefore likely to be more important than 
competitive-like accretion (ABMP07). However, given the total mass of 
the parent core, $5.9\pm1.3$ M$_{\sun}$, there is a considerable mass 
reservoir for (some of) the condensations to increase their mass via accretion, 
although by how much remains unclear. High-resolution spectral-line imaging of 
the whole core would be necessary to study whether there are 
infall motions towards the condensations and to estimate how much 
additional mass they could accrete (cf.~\cite{hacar2011}).

\subsection{Molecular column densities and fractional abundances}

To properly determine the column density of the molecules, we must 
first determine the optical thickness of the transition. The optical 
thickness of a spectral line can be derived 
from the radiative transfer equation, provided the line intensity and 
excitation temperature are known (as mentioned in Sect.~2, the optical 
thickness could not be derived through fitting the hf structure). 
We used the line peak intensities ($T_{\rm MB}$) to calculate the peak optical 
thicknesses, $\tau_0$ [see, e.g., Eq.~(A.3) in \cite{miettinen2012a}]. 
The C$^{17}$O$(2-1)$ transition was assumed to be thermalised at the adopted
gas temperature of the condensations, i.e., we set $T_{\rm ex}=10$ K. 
This assumption is supported by the results obtained in Paper III 
[$T_{\rm ex}({\rm C^{17}O})\simeq T_{\rm kin}$]. For the $J=3-2$ transition of 
N$_2$H$^+$ and N$_2$D$^+$, we adopted the value $T_{\rm ex}=5$ K. 
This is the typical value derived for the $J=3-2$ line or lower-$J$ lines 
of N$_2$H$^+$ and N$_2$D$^+$ (e.g., \cite{caselli2002}; \cite{crapsi2005}; 
Papers II--III; \cite{friesen2013}). Finally, the background temperature 
was assumed to be equal to the cosmic background radiation temperature of 
2.725 K. The derived $\tau_0$ values are listed in Col.~(7) of 
Table~\ref{table:lineparameters}, with the uncertainties propagated from those 
in $T_{\rm MB}$.

The beam-averaged C$^{17}$O column densities, $N({\rm C^{17}O})$, were derived
following the standard LTE analysis outlined, e.g., in the
paper by Miettinen (2012a; Appendix~A.3 therein). Because the $\tau_0$ values 
of the C$^{17}$O lines are so small, even the total optical thicknesses are 
$\tau_{\rm tot}\ll 1$ (as seen in a number of other studies; e.g., 
\cite{ladd1998}; \cite{fuller2002}). The $N({\rm C^{17}O})$ values were 
therefore computed from the integrated line intensities under the assumption 
of optically thin line emission. In the two cases where the detected line does 
not cover all the hf components (SMM 6a and 2nd velocity component towards 
SMM 6c), the column density was scaled by the inverse of the relative line 
strength within the detected line. The errors in $N({\rm C^{17}O})$ were 
propagated from those associated with the integrated intensity. 

Because the strongest hf component of the $J=3-2$ transition of N$_2$H$^+$ and 
N$_2$D$^+$ has a relative strength of only $11/63=0.1746$, the $\tau_{\rm tot}$ 
values inferred from $\tau_0$ are clearly above unity in most cases, i.e., 
the lines are optically thick (as is often the case for these lines; e.g., 
\cite{crapsi2004}; \cite{fontani2006}). The beam-averaged column densities of 
N$_2$H$^+$ and N$_2$D$^+$ were therefore computed from the FWHM linewidths and 
$\tau_{\rm tot}$ [see, e.g., Eq.~(10) in Paper II]. The column density 
uncertainties were derived from those of $\Delta {\rm v}$ and $\tau_{\rm tot}$. 
If we are overestimating the true linewidths, the 
corresponding column densities would represent upper limits in this sense. 
We also checked our derivation of the column density by examining 
some additional optically thin hf satellites. The N$_2$H$^+$ line towards 
SMM 6a offered the best opportunity to do this because in this case five hf 
lines lying 1.5--2.4 km~s$^{-1}$ from the strongest component were 
clearly detected (Fig.~\ref{figure:spectra}). The integrated intensity of the 
satellite group is $0.05\pm0.04$ K~km~s$^{-1}$ and its overall relatively 
strength is about 0.035. These can be converted into a total $N({\rm N_2H^+})$ 
of $1.5\pm1.2\times10^{13}$ cm$^{-2}$, which is $0.6\pm0.5$ times the value 
derived from $\tau_{\rm tot}$. The results are in reasonable agreement within 
the error.

The fractional abundances of the molecules were calculated by dividing
their column densities by the H$_2$ column density. For this purpose, the 
$N({\rm H_2})$ values were derived from the SABOCA map smoothed to the 
corresponding resolution of the line observations. The abundance errors
were derived by propagating the errors in the column densities of the observed 
species and H$_2$ molecules. The derived column densities and abundances are 
listed in Table~\ref{table:column}.

\subsection{CO depletion and deuterium fractionation}

To investigate the amount of CO depletion in the condensations, we 
calculated the CO depletion factors following the analysis presented in 
Sect.~4.5 of Paper III. In summary, the canonical (or undepleted)
CO abundance was adopted to be $9.5\times10^{-5}$ (\cite{frerking1982}). 
With the additional assumptions about the oxygen-isotopic ratios, namely 
$[{\rm ^{16}O}]/[{\rm ^{18}O}]=500$ and 
$[{\rm ^{18}O}]/[{\rm ^{17}O}]=3.52$, the canonical C$^{17}$O 
abundance was set to $5.4\times10^{-8}$. This value was then divided by the 
observed C$^{17}$O abundance, which gives the CO depletion factor, $f_{\rm D}$ 
(the eighth column of Table~\ref{table:column}). 
The $f_{\rm D}$ uncertainty was propagated from that in the observed abundance 
but it probably underestimates the true uncertainty by a factor of $2-3$ 
because of the uncertainties in the assumptions used (e.g., canonical 
abundance, oxygen-isotopic ratios, etc.). For instance, Lacey et al. (1994) 
found the best-fitting CO abundance of $2.7^{+6.4}_{-1.2}\times10^{-4}$
towards NGC 2024 in Orion B. Adopting this value would almost triple our 
$f_{\rm D}$ values. A lower limit to the condensation's age can be 
estimated to be comparable to its CO depletion timescale. Following the 
analysis presented in Miettinen (2012a, Sect.~5.5 therein), at a 
characteristic density of a few times $10^5$ cm$^{-3}$, the depletion time is 
$\tau_{\rm dep}\sim2.2\times10^4$ yr.

Towards each condensation, the N$_2$H$^+$ and N$_2$D$^+$ lines have 
similar radial velocities and linewidths. Therefore, the two
lines are probably tracing the same gas. It is therefore reasonable to 
calculate the degree of deuterium fractionation by dividing the column 
density of the deuterated isotopologue N$_2$D$^+$ by its normal 
hydrogen-bearing form N$_2$H$^+$. The error was derived from the errors
in the corresponding column densities. The results are shown in the last 
column of Table~\ref{table:column}.

\begin{table*}
\caption{Molecular column densities, fractional abundances with respect to 
H$_2$, CO depletion factors, and the deuteration degrees.}
\begin{minipage}{2\columnwidth}
\centering
\renewcommand{\footnoterule}{}
\label{table:column}
\begin{tabular}{c c c c c c c c c}
\hline\hline 
Source & $N({\rm C^{17}O})$ & $N({\rm N_2H^+})$ & $N({\rm N_2D^+})$ & $x({\rm C^{17}O})$ & $x({\rm N_2H^+})$ & $x({\rm N_2D^+})$ & $f_{\rm D}$ & $N({\rm N_2D^+})/N({\rm N_2H^+})$\\
       & [$10^{14}$ cm$^{-2}$] & [$10^{13}$ cm$^{-2}$] & [$10^{12}$ cm$^{-2}$] & 
[$10^{-8}$] & [$10^{-9}$] & [$10^{-10}$] & & \\
\hline
SMM 6a & $1.1\pm0.2$ & $2.5\pm0.5$\tablefootmark{a} & $7.6\pm0.6$ & $1.5\pm0.6$ & $2.3\pm0.9$ & $9.9\pm3.5$ & $3.6\pm1.5$ & $0.30\pm0.07$ \\  
SMM 6b & $2.5\pm0.3$ & $1.2\pm0.2$ & $5.2\pm0.7$ & $3.5\pm1.4$ & $1.3\pm0.5$  & $7.2\pm2.8$ & $1.5\pm0.6$ & $0.43\pm0.09$ \\
SMM 6c & $3.0\pm0.3$ & $0.7\pm0.1$ & $2.9\pm0.2$ & $5.4\pm2.2$ & $0.9\pm0.4$  & $4.9\pm1.9$ & $1.0\pm0.4$ & $0.41\pm0.07$ \\
SMM 6c (2nd v-comp.) & $0.5\pm0.3$ & $0.3\pm0.1$ & \ldots & $0.9\pm0.6$ & $ 0.4\pm0.1$  &  \ldots & $5.9\pm4.1$ & \ldots\\
SMM 6d & $2.7\pm0.3$ & $0.7\pm0.1$ & $2.5\pm0.8$ & $6.4\pm3.1$ & $1.1\pm0.5$ & $5.5\pm3.0$ & $0.8\pm0.4$ & $0.36\pm0.13$\\
\hline 
\end{tabular} 
\tablefoot{The $N({\rm C^{17}O})$ values were computed by assuming that 
$T_{\rm ex}=10$ K. The N$_2$H$^+$ and N$_2$D$^+$ column densities were derived 
under the assumption that $T_{\rm ex}=5$ K.\tablefoottext{a}{The value 
derived from a group of five optically thin satellites is 
$1.5\pm1.2\times10^{13}$ cm$^{-2}$ (see text).}}
\end{minipage} 
\end{table*}







\section{Discussion}

\subsection{Substructure within Ori B9--SMM 6}

The studied core SMM 6 is one of the rare examples of a prestellar core 
that is fragmented into smaller condensations. Even though the theoretical 
predictions of the cylindrical fragmentation are not 
exactly similar to the observed characteristics, it still seems likely that 
the core has fragmented as a result of Jeans-type gravitational instability. 
In particular, the projected separation between the fragments is very close to 
the local thermal Jeans length. The radial crossing time of the filamentary 
parent core suggests that its fragmentation timescale is $\sim0.18$ Myr.

The estimated masses of the condensations are $\sim0.2-0.6$ M$_{\sun}$ 
with a typical uncertainty of $\sim0.1$ M$_{\sun}$. The condensations are 
characterised by subsonic non-thermal motions with 
$\sigma_{\rm NT}\simeq 0.5c_{\rm s}$ assuming that the characteristic gas 
temperature is 10 K. The calculated condenstation virial parameters 
have relatively large errors, but it seems likely that they 
are self-gravitating. The minimum mass for a main-sequence star is 
$M_{\star}=0.08$ M$_{\sun}$, below which the stellar temperature is not high 
enough to ignite proton-proton nuclear fusion. If the star formation 
efficiency for individual condensations is comparable to that of dense cores, 
$\epsilon_{\rm core}=M_{\star}/M_{\rm core}\sim30-50\%$ 
(e.g., \cite{matzner2000}; \cite{goodwin2008}; \cite{andre2009}), they might be 
able to collapse to form stars without any additional mass accretion. 
At least for the most massive fragment, SMM 6a, this seems possible. 
The lowest mass fragments could also form substellar-mass objects or brown 
dwarfs ($0.012<M/{\rm M_{\sun}}<0.08$; e.g., \cite{molliere2012}). 
Examples of brown dwarfs that were likely formed or are currently 
forming via direct collapse include the wide (800 AU) binary brown dwarf 
system FU Tau A/B (\cite{luhman2009}) and the pre-brown dwarf condensation 
Oph B-11 discovered by Andr\'e et al. (2012). Some of the very low luminosity 
objects, such as L1148-IRS, could also be the precursors of isolated brown 
dwarfs (\cite{kauffmann2011}). Brown dwarfs may also form (and 
subsequently be ejected) through disk fragmentation (see, e.g., 
\cite{stamatellos2009}). In theory, the minimum mass of an object capable of 
forming through gas fragmentation is set by the opacity limit for radiative 
cooling and is in the range $\sim0.001-0.004$ M$_{\sun}$, that is to say, 
fragmentation could even form planetary-mass objects 
(e.g., \cite{whitworth2006}).

The dispersion of the condensation velocity centroids is 
smaller than the one-dimensional velocity dispersion of the 
parent core gas by a factor of 11 (0.02 vs. 0.22 
km~s$^{-1}$, where the latter is derived from $c_{\rm s}$ at 11 K and 
0.1 km~s$^{-1}$ non-thermal contribution). Synthetic observations 
(e.g., N$_2$H$^+$) of simulated environments (decaying and driven turbulence) 
by Offner et al. (2008) also show that on the scale of cores, the dispersion 
of the velocity centroids is small (cf.~\cite{offner2009}). 
Indeed, the motions of the condensations relative to one another appear to be 
so slow that they have no time to coalesce or gather together before 
evolving individually into protostars or proto-brown dwarfs. Moreover, 
despite the relatively large mass reservoir ($\sim 6$ M$_{\sun}$) 
surrounding the condensations, they are unlikely to be able to increase their 
mass via competitive-like accretion. Instead, the fragments' internal 
self-gravity could induce inward motions and accretion of additional 
mass -- a scenario that could be tested by future high-resolution 
observations.

\subsection{CO depletion and deuterium fractionation in the 
condensations}

The gas phase depletion factors of CO derived towards the condensations are 
in the range $f_{\rm D}=0.8\pm0.4-3.6\pm1.5$. Therefore, there is no 
evidence of significant CO depletion. There are a few examples of high-density 
($\sim 10^5$ cm$^{-3}$) cores where CO depletion is not prominent,
namely L1495B, L1521B, and L1521E (\cite{tafalla2004}; \cite{hirota2004}). 
Deuterium fractionation in these Taurus cores is also found to be low 
(\cite{hirota2004}, and references therein), and they are considered to be 
both chemically and dynamically young.

However, as mentioned in Sect.~4.6, the undepleted CO abundance in 
Orion B9 might well be higher than what we have adopted in the analysis. 
For this reason, we might be underestimating the $f_{\rm D}$ values by 
a factor of about three (\cite{lacey1994}). Also, we have observed the $J=2-1$ 
rotational transition of C$^{17}$O, the line whose 10-K critical density is 
$2.1\times10^4$ cm$^{-3}$. Apart from the condensations, we might therefore be 
tracing lower density gas along the line of sight where CO is not 
significantly depleted (e.g., \cite{fontani2012}). 
The $27\farcs8$ resolution of our C$^{17}$O observations also provides us 
with only beam-averaged $f_{\rm D}$ values. The above factors are likely to 
explain the $f_{\rm D}$ value less than unity derived for SMM 6d. 
Indeed, the temperatures and densities of the condensations are likely to be 
so low and high, respectively, that significant CO freeze-out is expected 
(e.g., \cite{leger1983}; \cite{rawlings1992}). 
For example, Tafalla et al. (2004) estimated 
that the density at which CO disappears from the gas phase is only 
$7.8\times10^4$ cm$^{-3}$ and $2.5\times10^4$ cm$^{-3}$ for the starless cores 
L1498 and L1517B in Taurus, respectively. Moreover, for our condensations the 
CO freeze-out timescale is estimated to be shorter than the free-fall time by 
a factor of about four (see \cite{bergin2007}) and about eight times shorter 
than the fragmentation timescale ($\tau_{\rm cross}$). 

For comparison, Caselli et al. (1999) found that CO is depleted by a factor of 
$\sim10$ at the dust peak of the prestellar core L1544. Similarly, 
Savva et al. (2003) found that, relative to the canonical abundances from 
Frerking et al. (1982), CO is depleted by a factor of $\sim10$ in a sample 
of nine dense cores in Orion B. More recently, Christie et al. (2012) derived 
the mean (median) CO depletion factor of 19 (10) towards starless cores in the 
NGC 2024 region of Orion B. We note that in Paper III, where we also employed 
APEX/C$^{17}$O$(2-1)$ observations, we derived the value 
$f_{\rm D}=4.2\pm1.3$ towards a position lying $11\arcsec$ north of the 
350-$\mu$m peak of SMM 6a (where we now obtain the value $3.6\pm1.5$). 
One would expect to see stronger depletion towards the denser dust peak, but 
there are two reasons that explain this discrepancy: \textit{i}) the C$^{17}$O 
column densities in Paper III were derived through non-LTE RADEX analysis 
(\cite{vandertak2007}), and \textit{ii}) the H$_2$ column densities were 
computed from the LABOCA 870-$\mu$m peak flux densities assuming a thick ice 
mantle dust model of OH94 and a dust-to-gas ratio of 1/100 (see 
Appendix~A). 

When CO molecules freeze out on the surface of dust grains, the rate
of gas-phase deuterium fractionation increases (e.g., \cite{dalgarno1984}). 
The reason for this is that the main destruction path of H$_2$D$^+$, namely 
${\rm H_2D}^++{\rm CO}\rightarrow {\rm HCO}^+\,{\rm or}\,{\rm DCO}^+$, is no 
longer effective, enabling H$_2$D$^+$ to donate its deuteron to other species 
(e.g., ${\rm H_2D}^++{\rm N_2}\rightarrow {\rm N_2H}^+\,{\rm or}\,{\rm N_2D}^+$).
The levels of N$_2$D$^+$/N$_2$H$^+$ deuteration we derive here, 
$0.30\pm0.07-0.43\pm0.09$, are relatively high compared to those found in 
other studies of starless cores. For example, Crapsi et al. (2005) found 
N$_2$D$^+$/N$_2$H$^+$ column density ratios in the range $<0.02-0.44\pm0.08$ 
for their sample of starless/prestellar cores. Friesen et al. (2013) derived 
the values $\lesssim0.02-0.20\pm0.04$ towards starless and protostellar cores 
in Perseus with no significant difference between the two types of objects.
Through radiative transfer modelling of the starless core L183, 
Pagani et al. (2007) derived a deuteration level of $0.70\pm0.12$ at the 
CO-depleted core centre, exceeding the highest value found in the present study.
As for $f_{\rm D}$, a high value of deuteration ($0.60\pm0.08$) was derived in 
Paper III towards the edge of SMM 6. Again, the column density calculations 
were based on RADEX analysis (N$_2$H$^+$) and LTE modelling with CLASS/Weeds 
(N$_2$D$^+$), and are therefore not directly comparable with the present 
results. 

The deuterium fractionation of N$_2$H$^+$ and the amount of CO depletion are 
found to be positively correlated in starless/prestellar cores and the 
envelopes of Class 0 protostars (e.g, \cite{crapsi2004}, 2005; 
\cite{emprechtinger2009}). This is not evident among our condensations. 
In fact, the most CO-depleted fragment, SMM 6a, shows the lowest level of 
deuteration. Our source sample is, however, very small, and within the error 
bars the values of $f_{\rm D}$ and $N({\rm N_2D^+})/N({\rm N_2H^+})$ are 
comparable to each other, suggesting that they are in a similar stage of 
chemical and dynamical evolution. This conforms to the scenario where the 
condensations were formed simultaneously via core fragmentation. The high 
levels of molecular D/H ratio suggest that they are highly evolved, and 
possibly at the onset of protostellar collapse. This seems to 
contradict the low depletion factors in the condensations, but as discussed 
above we are very likely underestimating the $f_{\rm D}$ values.

\section{Summary and conclusions}
   
We have used the APEX telescope to observe the molecular-line transitions 
C$^{17}$O$(2-1)$, N$_2$H$^+(3-2)$, and N$_2$D$^+(3-2)$ towards the low-mass 
condensations inside the prestellar core Ori B9--SMM 6 which were discovered 
by our previous SABOCA 350-$\mu$m imaging of the core (Paper III). 
The present work is one of the first studies of the dynamics of subfragments 
within a prestellar core.

The condensations were likely formed as a result of Jeans-type fragmentation 
of the parent core. Based on the C$^{17}$O velo\-city dispersion data, we 
estimate that the fragmentation timescale of the filamentary parent core 
is $\sim1.8\times10^5$ yr. The condensations show only subsonic internal 
non-thermal motions ($\sigma_{\rm NT}=0.1$ km~s$^{-1}$ if $T_{\rm kin}=10$ K), 
and most of them are likely to be gravitationally bound. 
The condensation-to-condensation kinematics was investigated by employing the 
velocity information provided by the high-density tracer N$_2$H$^+(3-2)$. 
The estimated timescale required for the condensations to coalesce is a few 
Myr, which is much longer than their free-fall timescale. It is therefore 
unlikely that the subfragments will have time to coalesce before collapsing 
into protostars or proto-brown dwarfs, or fragment even further down to 
planetary-mass objects. Significant accretion in a competitive fashion
between the condensations from the parent core's mass reservoir is also 
unlikely. The reason for this is that the condensations' self-gravity is 
expected to lead to a much higher (by a factor of $\sim60$) mass infall 
rate than what is provided by competitive-like process. 

The present molecular data were also used to study the amount of CO depletion 
and deuterium fractionation in the condensations. The CO depletion factors we 
derive, $f_{\rm D}=0.8\pm0.4-3.6\pm1.5$, do not suggest any significant CO 
freeze-out. However, these values are based on the canonical CO abundance of 
$9.5\times10^{-5}$ relative to H$_2$, whereas in Orion B this value can be 
considerably higher (by a factor of $\sim3$). Therefore, at least an 
intermediate level of CO depletion seems possible in the condensations. 
The CO depletion timescale for the condensations ($\sim2.2\times10^4$ yr) is 
shorter than the fragmentation timescale (by a factor of $\sim8$) and the 
free-fall time of the condensations ($\sim 7\times10^4$ yr). 
Higher-resolution observations would undoubtedly reveal more CO-depleted 
condensation interiors. The derived N$_2$H$^+$ and N$_2$D$^+$ column densities 
lie in the range $0.7-2.5\times10^{13}$ and $2.5-7.6\times10^{12}$ cm$^{-2}$, 
and the corresponding fractional abundances are 
$\sim0.9-2.3\times10^{-9}$ and $\sim4.9-9.9\times10^{-10}$, 
respectively. The deuterium fractionation of 
N$_2$H$^+$, or the N$_2$D$^+$/N$_2$H$^+$ column density ratio, lies in the range 
$0.30\pm0.07-0.43\pm0.09$. This is a stronger level of deuteration 
than is typically observed in starless/prestellar cores and should 
require significant gas-phase depletion of CO. The very high molecular 
deuteration observed towards the condensations suggests that they are in an 
advanced stage of chemical evolution and possibly on the verge of 
gravitational collapse. The actual fate of the subfragments, i.e., whether 
they can collapse into protostellar or substellar objects, remains to be 
elucidated. The lowest-mass condensations within SMM 6 could be the precursor 
sites of brown dwarf formation.

\begin{acknowledgements}
    
We are grateful to the staff at the APEX telescope for performing the 
service-mode heterodyne observations presented in this paper. We also 
appreciate the positive comments and insight of the anonymous referee. O.~M. 
acknowledges the Academy of Finland for the financial support through grant 
132291. S.~S.~R.~O. acknowledges support from NASA through Hubble Fellowship 
grant HF-51311. This research has made use of NASA's Astrophysics Data System 
and the NASA/IPAC Infrared Science Archive, which is operated by the JPL, 
California Institute of Technology, under contract with the NASA. SPIRE has 
been developed by a consortium of institutes led by Cardiff Univ. (UK) and 
including: Univ. Lethbridge (Canada); NAOC (China); 
CEA, LAM (France); IFSI, Univ. Padua (Italy); IAC (Spain); Stockholm 
Observatory (Sweden); Imperial College London, RAL, UCLMSSL, UKATC, Univ. 
Sussex (UK); and Caltech, JPL, NHSC, Univ. Colorado (USA). This development 
has been supported by national funding agencies: CSA (Canada); NAOC (China); 
CEA, CNES, CNRS (France); ASI (Italy); MCINN (Spain); SNSB (Sweden); STFC, 
UKSA (UK); and NASA (USA).
  
\end{acknowledgements}

\begin{appendix} 
\section{Revision of core/condensation properties presented in 
Paper III}

In Paper III, the condensation masses and densities were estimated 
assuming that $T_{\rm dust}$ equaled the $T_{\rm kin}$ derived towards 
the selected position near the edge of SMM 6, i.e., $11.0\pm0.4$ K. However, 
all the subcondensations except SMM 6a lie outside the $40\arcsec$ beam of 
the NH$_3$ measurements used to derive $T_{\rm kin}$. Moreover, $T_{\rm dust}$ 
is expected to be lower in the embedded small condensations because they 
are better shielded from the external dust heating radiation field. On
these grounds, we revise the temperature-dependent condensation properties 
by assuming that $T_{\rm dust}=10$ K. This seems a reasonable choice given 
that the gas temperature is about 11 K at the core edge. 

In addition, it was previously assumed that the 350-$\mu$m dust mass 
absorption (or emission) coefficient, i.e., the dust opacity per unit 
dust mass, was $\kappa_{{\rm 350\, \mu m}}=10$ cm$^2$~g$^{-1}$. 
This value was interpolated from the widely used Ossenkopf \& Henning (1994, 
hereafter OH94) model describing graphite-silicate dust grains that have 
coagulated and accreted \textit{thick} ice mantles over a period of $10^5$ yr 
at a gas density of $n_{\rm H}=10^5$ cm$^{-3}$. If we assume that grains have 
\textit{thin} ice mantles, which might be more appropriate here (e.g., 
\cite{stamatellos2003} and references therein)\footnote{The assumption of 
thin ice mantles is supported by the fact that no significant CO depletion is 
found in the present study (see Sect.~4.6).}, 
$\kappa_{{\rm 350\, \mu m}}$ is decreased to 7.84 cm$^2$~g$^{-1}$. 
In Paper III, the dust-to-gas ratio was assumed to be 1/100. However, 
this value refers to the canonical dust-to-hydrogen mass ratio, 
$M_{\rm dust}/M_{\rm H}$ (e.g., \cite{draine2011}; Table~23.1 therein). If 
we assume solar composition, i.e., the mass fractions for hydrogen, helium, 
and heavier elements are $X=0.71$, $Y=0.27$, and $Z=0.02$, respectively, 
the ratio of total mass (H+He+metals) to hydrogen mass is $1/X\simeq1.41$. 
The total dust-to-gas mass ratio is therefore 
$M_{\rm dust}/M_{\rm gas}=M_{\rm dust}/(1.41M_{\rm H})=1/141$. We note that for the 
assumed gas composition, the mean molecular weight per free particle is 
$\mu_{\rm p}=2.37$ and that per H$_2$ molecule is $\mu_{\rm H_2}\simeq2.82$ 
(\cite{kauffmann2008}; Appendix~A.1 therein).

The masses ($M$), peak beam-averaged H$_2$ column densities [$N({\rm H_2})$], 
and averaged H$_2$ number densities [$\langle n({\rm H_2}) \rangle$] were 
computed using the standard optically thin dust emission formulation 
[see, e.g., Eqs.~(2) and (3) in Paper I, and Eq.~(1) in Paper III]. 
The results are given in Table~\ref{table:sources}. 
The uncertainties in $M$ and $N({\rm H_2})$ were propagated from the 
uncertainties in the integrated flux density and the peak surface
brightness, respectively. We note that the uncertainty in dust opacity, which 
is likely a factor $\gtrsim2$, is the major source of error in the mass 
and column density estimate. The error in $\langle n({\rm H_2}) \rangle$ was 
propagated from that of $M$. 

We note that the total mass of the SMM 6 core was previously determined 
from the LABOCA 870-$\mu$m emission adopting the thick ice mantle model of 
OH94, in which case $\kappa_{{\rm 870\, \mu m}}\simeq1.7$ cm$^2$~g$^{-1}$. 
If we employ the above thin ice mantle model, $\kappa_{{\rm 870\, \mu m}}$ is 
1.38 cm$^2$~g$^{-1}$. Moreover, the 870-$\mu$m flux density of the source, when 
integrated inside the 0.09 Jy~beam$^{-1}$ or $3\sigma$ contour of the 
re-reduced LABOCA map (see \cite{miettinen2013}), is 
$S_{{\rm 870\, \mu m}}=1.03\pm0.22$ Jy. For $T_{\rm dust}=11.0\pm0.4$ K 
and $M_{\rm dust}/M_{\rm gas}=1/141$, these 
revisions yield a mass of $5.9\pm1.3$ M$_{\sun}$, a factor of 
1.4 lower than reported previously (see Paper II). The 
discrepancy mostly stems from the larger area previously used to compute 
$S_{{\rm 870\, \mu m}}$. The virial parameter of the core from Paper II is 
accordingly revised to $\alpha_{\rm vir}=1.1\pm0.2$, which still 
satisfies the gravitational boundedness condition of $\alpha_{\rm vir}<2$ 
and is very close to virial equilibrium ($\alpha_{\rm vir}=1$). 
Therefore, the source can be considered a candidate prestellar core 
(cf.~Sect.~4.2).

The projected length of the SMM 6 filament along its long 
axis, as measured from the \textit{Herschel} 250-$\mu$m emission shown in 
Fig.~\ref{figure:SMM6}, is about $1\farcm7$ or $L=0.22$ pc. 
This corresponds to the extent of the $3\sigma$ LABOCA 870-$\mu$m emission 
(Paper III; Fig.~9 therein). The corresponding mass per unit length, or line 
mass, is $M_{\rm line}\simeq 27\pm6$ M$_{\sun}$~pc$^{-1}$.

\end{appendix}

\end{document}